\documentclass[twocolumn,letterpaper,aps,prc,superscriptaddress,nofootinbib,showpacs,floatfix]{revtex4-1}
\usepackage{graphicx}
\usepackage{comment}
\usepackage{setspace}
\usepackage{amsmath}
\usepackage{amssymb}
\usepackage[textwidth=16cm,textheight=22cm]{geometry}
\usepackage{color}
\usepackage{indentfirst}
\usepackage{xspace}
\usepackage{hyperref}
\usepackage{verbatim}
\usepackage{epstopdf}
\usepackage{csquotes}
\usepackage{multirow}
\usepackage[utf8]{inputenc}

\usepackage{lineno}
{\large }

\newcommand {\pbpb} {\mbox{Pb$-$Pb}~}
\newcommand {\rootsNN} {\mbox{$\sqrt{s_{\rm NN}}$}}

\begin{document}

\title{Identical-particle (pion and kaon) femtoscopy in \pbpb collisions at \rootsNN = 5.02 TeV with Therminator2 modeled with (3+1)D viscous hydrodynamics}
\author{Pritam Chakraborty} 
\email{prchakra@iitb.ac.in}
\affiliation{Indian Institute of Technology Bombay, Mumbai, India}
\author{Ashutosh Kumar Pandey}
\email{ashutosh.kumar.pandey@cern.ch}
\affiliation{Indian Institute of Technology Bombay, Mumbai, India}
\affiliation{University of Tsukuba, Japan}
\author{Sadhana Dash}
\email{sadhana@phy.iitb.ac.in}
\affiliation{Indian Institute of Technology Bombay, Mumbai, India}


\date{\today}  


\begin{abstract}
The three-dimensional femtoscopic correlations of pions and kaons are presented for Pb$-$Pb collisions at  \rootsNN = 5.02 TeV within the framework of 
 (3+1)D viscous hydrodynamics  combined with THERMINATOR 2 code for statistical hadronization. The femtoscopic radii for pions and kaons are obtained as a function 
 of  pair transverse momentum and centrality in all three pair directions. The radii showed a decreasing trend with an increase of pair transverse momentum 
 and transverse mass for all centralities. These observations indicate the presence of strong collectivity. A simple effective scaling of radii with pair 
 transverse mass was observed for both pion and kaons.  
 
\end{abstract}

\maketitle

\section{Introduction}
The novel phase of matter characterised by the existence of deconfined state of quarks and gluons is expected to be created at relativistic heavy ion collisions 
at  RHIC and LHC energies \cite{starwhite, phenixwhite}.  The initial state of the matter is well described by the equations of hydrodynamics and the hydrodynamic collectivity is further ascertained through the measurement of observables like  transverse momentum  spectra of measured particles, radial and elliptic flow and  two-particle angular 
correlations. However, most of the observables were momentum-based observables and very little was known about the effects of collectivity on spatio-temporal  characteristics of
particle production and correlation. It is necessary to know the space-time dimension of the system produced in heavy-ion collisions to understand  the hydrodynamic collectivity 
\cite{sinyukov, bronio}. 
The extremely small size and lifetime of the system created make it impossible to measure its dimensions directly by any other means. Femtoscopy 
provides a unique probe to extract the source size of the system by means of studying two particle correlations in momentum space \cite{mikelisa}. It 
incorporates Bose-Einstein (Fermi-Diarc) correlation between two identical particles due to symmetrization (anti-symmetrization) of wave functions of 
identical bosons (fermions) and Final state interaction (FSI). The analysis is performed in the Longitudinally Comoving System (LCMS), where the pair 
longitudinal (\enquote{long}) axis is along the beam direction, the outward (\enquote{out}) axis is along the transverse momentum of the pair and the \enquote{side} axis is 
perpendicular to the other two axes. In the LCMS, the total momentum of the pairs along the longitudinal direction is conserved. The sizes or the \enquote{femtoscopic 
radii} of the system along the $out$, $side$ and $long$ directions can be expressed as $R_{out}$, $R_{side}$ and $R_{long}$, respectively which can be
calculated as function of pair transverse momentum and event centrality. The previous measurements at RHIC, BNL at 200 GeV and at LHC, CERN at 2.76 TeV 
revealed that the system size decreases as the pair transverse momentum ($k_T$) increases and also, it decreases from central to peripheral events. Besides, 
at both RHIC and LHC energies, the radii extracted for the pions showed  a power-law dependence (scaling) on transverse mass ($m_T$) and a linear dependence on the 
final state multiplicity $(\frac{dN_{ch}}{d\eta})^\frac{1}{3}$ \cite{star1,star2,phenix1,alicepion}. The findings were consistent with the existence of hydrodynamic collectivity.  
One expects the scaling of femtoscopic radii to be extended to other particle species (such as kaons, protons etc) as they experience the same flow field which 
the pions were subjected to.  A detailed study of the identical particle femtoscopic studies using the (3+1)D viscous hydrodynamics coupled with statistical hadronization THERMINATOR 2 model \cite{thermi1,thermi2} extended the study to 
kaons and pions and observed an approximate scaling of $m_{T}$ for pions, kaons and protons in Pb$-$Pb collisions at \rootsNN = 2.76 TeV \cite{Therminator:2760MeV}. 
However,  hadronic re-scattering, resonance decays and other final state interactions can significantly affect the observed scaling. Recent results from PHENIX \cite{phenix2} and ALICE \cite{alicekaon} experiments have reported the breaking of the $m_{T}$ scaling in kaon-kaon correlations while it is not broken for $k_{T}$. 
The hydrokinetic model (HKM) \cite{hkm1,hkm2} which includes the hydrodynamic phase and a hadronic re-scattering phase  described the femtoscopic radii for pions and kaons and predicted the violation of  $m_{T}$ scaling between pion and kaons at LHC enrgies \cite{hkm3}. The work showed that the approximate scaling was restored when the re-scattering phase was not considered. This indicates the importance of the influence of hadronic re-scattering phase.

\par In this work, a femtoscopic study of identical pions and kaons have been performed using (3+1D) viscous hydrodynamics coupled with THERMINATOR 
2 model at \rootsNN = 5.02 TeV and the $m_T$ and $(\frac{dN_{ch}}{d\eta})^{1/3}$ scaling of the femtoscopic radii have also been reported. The aim of the work is to 
verify whether the change in freeze-out temperature affects the approximate $m_{T}$ scaling observed previously at \rootsNN = 2.76 TeV and to provide the initial 
predictions on the source radii at freeze-out  using  pion and kaon femtoscopy.
Due to the statistical limitations, the femtoscopic calculations for protons have been performed in one dimension  together with pions and kaons in 
the Pair Rest Frame (PRF). The one-dimensional radii for pions, kaons and protons were extracted  and  the relation between the femtoscopic radii in LCMS and in PRF 
was explored.

\label{intro}
\section{Initial hydrodynamic conditions with Therminator 2}
THERMINATOR 2 is a Monte Carlo event generator written in C++ with the standard 
CERN ROOT environment. 
In this work, the events were generated using the Therminator 2 model which combines the (3 + 1)D viscous hydrodynamics and 
the statistical hadronisation together with resonance propagation and decay of unstable particles.
\par The evolution of the flow velocity $u^\mu$ and energy density
$\epsilon$ of the system are solved in  $(3+1)$D  viscous hydrodynamics model using the second-order Israel-Stewart 
equations \cite{Israel-Stewart}.The energy-momentum tensor, $T^{\mu\nu}$ consists of ideal fluid part, stress tensor 
$\pi^{\mu\nu}$ and bulk viscosity correction $\Pi$ \cite{Therminator:2760MeV}.
The shear viscosity to entropy density, $\eta/s =0.08$ was used while bulk viscosity to entropy density, was kept at  $\zeta/s=0.04$.  The initial time, $\tau_{i}$ 
for the hydrodynamic evolution was 0.6 fm$/c$  and the freeze-out temperature is $T_f = 150$ MeV. The chemical potentials were all set to zero.
\begin{eqnarray}
 T^{\mu\nu}=(\epsilon+p+\Pi)u^{\mu}u^{\nu}+\pi^{\mu\nu}-(p+\Pi)g^{\mu\nu}
\end{eqnarray}

\par After hydrodynamic evolution of the system, the event-by-event fluctuations
in the initial conditions produce fluctuating freeze-out hyper-surfaces. The 
Glauber model was used to  obtain the initial density profile in the transverse 
plane, $\rho_{part}\frac{1-\alpha}{2}+\alpha\rho_{bin}$, where $\rho_{bin}$
and $\rho_{part}$ are the density of binary collisions and participating 
nucleons, respectively and the mixing parameter, $\alpha=0.15$ \cite{Therminator:2760MeV}.  
This analysis has been done for seven sets of impact parameters values (in fm) for Pb-Pb collisions at the
$\sqrt{s_{NN}}=5.02$ TeV. The impact parameters of  2.3, 5.7, 7.4, 8.7, 9.9, 10.9 and 11.9  ( in fm), corresponds 
to the collision events at LHC of different centrality ranges: 0-5\%, 10-20\%, 20-30\%, 
30-40\%, 40-50\%, 50-60\% and 60-70\%, respectively \cite{Therminator:2760MeV, theory6} .
The yield of the particles, produced after hadronisation 
are calculated from the Cooper-Frye formula by THERMINATOR 2 code which uses the 
freeze-out hypersurfaces obtained from the hydrodynamic calculations as the input \cite{Therminator:2760MeV}.  
\begin{eqnarray}
 E\frac{d^3N}{dp^3}=\int d\Sigma_{\mu} p^{\mu}f(p_{\mu}u^{\mu})
\end{eqnarray}
where $d\Sigma_{\mu}$ is the integration element on the freeze-out hypersurface and
$f$ is the momentum distribution which includes non-equilibrium corrections to the bulk
viscosity $(\delta f_{bulk})$ and shear viscosity $(\delta f_{shear})$. 
The details of the modifications of the equilibrium momentum distributions and  
corrections can be found in \cite{Therminator:2760MeV}:
\par 
The THERMINATOR 2 model is based on single freeze-out and there is no difference between the chemical and kinetic freeze-out 
of the system. This model includes all known resonances in hadronisation along with their propagation
and decay. The particles are assumed to be created either on the freeze-out hypersurface
or from the subsequent decay of unstable particles. 

\section{Identical-particle femtoscopy methodology}
The femtoscopy method relies on two particle correlations originating primarily from wave function (anti)symmetrization 
commonly known as the Quantum Statistics (QS) effect. In this work, the formalism outlined in reference \cite{Therminator:2760MeV} has 
been followed to obtain the correlation function. 
The correlation function is defined as the ratio of probability of observing two particles with different momentum 
simultaneously to the  probabilities of observing each of them separately. It can be constructed by finding the  
ratio of the relative momentum distribution of particle pairs produced in same event
to the distribution of particle pairs selected from different events. The other sources of correlation namely 
the final state interactions due to Coulomb and Strong interaction between the particle pairs is not included. The effect of
QS is incorporated as an afterburner and is considered to be the only source of correlation in this analysis.
If $\Psi$ is the pair wave-function, it has to be symmetrized for the identical bosons e.g. pions,
 kaons and can be expressed as \cite{Therminator:2760MeV}
 
 \begin{eqnarray}
  \Psi_{K.\pi}=1+cos(2\bf{k^*r^*}), \label{eq:pairboson}
 \end{eqnarray}

and for the identical fermions e.g. protons, it can be expressed as \cite{Therminator:2760MeV}
 
 \begin{eqnarray}
  \Psi_{p}=1-\frac{1}{2}cos(2\bf{k^*r^*}), \label{eq:pairfermion}.
 \end{eqnarray} 
 
Theoretically, one can express the correlation function, $C( \bf {k^{*}})$ as \cite{Therminator:2760MeV}

\begin{eqnarray}
 C(\bf{k^*})=\frac{\int S(\bf{r^*},\bf{k^*})|\Psi(\bf{r^*},\bf{k^*})|^2}
 {\int S(\bf{r^*},\bf{k^*})} \label{eq:corr}
\end{eqnarray}

 where $\bf{r^*}$ is the relative space-time separation of two particles at the time of generation, 
 $\bf{k^*}$ is half of the pair relative momentum i.e. momentum of
 the first particle at Pair Rest Frame (PRF) and S is the source function which 
 corresponds to the probability of the emission of pair of particles. $\Psi$ is the 
 pair wave-function and hence accounts for the mutual interactions between the particles in the pair. 
 
 The correlation function has been constructed by filling two histograms, namely {\bf Num} and {\bf Den} 
 for pairs of identical particles for a given centrality class. The charged pions or kaons generated
 from Therminator 2 events were combined to form pairs. The {\bf Num} histogram was filled with particle pairs 
 corresponding to relative momenta ${\bf{q}}=2\bf{k^*}$ from same events. The pair weight for a particular $k_{T}$ bin was 
 calculated from Eq.(\ref{eq:pairboson}) for pions and kaons and from Eq.(\ref{eq:pairfermion}) for protons.
 The {\bf Den} histogram was filled with pairs where a particle is selected from mixed events and the weight given 
 while filling was 1.0. These histograms can be expressed as a function of
 three components of $\bf{q}$ (three dimensional),  $|\bf{q}|$(one 
 dimensional) only, or as a set of one-dimensional histograms corresponding to spherical
 harmonics representation of the pair distribution. 
 The correlation function was calculated as $C=Num/Den$. 
 \par In order to estimate the femtoscopic radii of the system, the source function was assumed to be a 
 three-dimensional ellipsoid with Gaussian density profile \cite{Therminator:2760MeV}:
 
 \begin{eqnarray}
  S({\bf{r}})\approx exp\left(-\frac{r^2_{out}}{4R^2_{out}}-\frac{r^2_{side}}{4R^2_{side}}-
  \frac{r^2_{long}}{4R^2_{long}}\right)  \label{source}
 \end{eqnarray}
 
 where $r_{out}$, $r_{side}$ and $r_{long}$ were the components of $\bf{r^*}$ calculated 
 in LCMS and $R_{out}$, $R_{side}$ and $R_{long}$ correspond to single-particle femtoscopic source
radii, in transverse, side and longitudinal directions, respectively. 

The radii in different directions were extracted by fitting the following function to the correlation function \cite{Therminator:2760MeV} obtained 
from Eq.(\ref{eq:corr}):

\begin{eqnarray} \nonumber
    C({\bf{q}})=1+\lambda exp ( -R^2_{out}q^2_{out}-R^2_{side}q^2_{side} \\
 -R^2_{long}q^2_{long}),  \label{fit3d}
\end{eqnarray}
 
where, $\lambda$  accounts for the strength of the correlation which signifies that not all pairs
considered were correlated and the assumed functional form of the source was not exactly Gaussian Eq.(\ref{eq:corr}).

\par Similarly,  the functional form of the source for one-dimensional correlation function was assumed to be spherically
symmetric in PRF and can be expressed as \cite{Therminator:2760MeV}:

\begin{eqnarray}
 S({\bf{r^*}})\approx exp\left(-\frac{\bf{{r}^*}^2}{4R_{inv}^2} \right), \label{eq:source1d}
\end{eqnarray}

where $R_{inv}$ is the source size.\\
The one-dimensional fitting function was \cite{Therminator:2760MeV}:

\begin{eqnarray}
 C(q_{inv})=1+\lambda exp \left( -R^2_{inv}q^2_{inv}\right).   \label{eq:fit1d}
\end{eqnarray}

The transformation from LCMS to PRF was performed by a Lorentz boost along
the pair transverse momentum with velocity $\beta_T = \frac{p_T}{m_T}$.  This also indicates that 
in the PRF, the $R_{side}$ and $R_{long}$ remain unchanged and only the $R_{out}$
changes according to the Eq. \ref{eq:LCMS_to_PRF}.
\begin{eqnarray}
R^*_{out}=\gamma_T R_{out}   \label{eq:LCMS_to_PRF}
\end{eqnarray}
where, the $`` * " $ corresponds to the quantity in PRF and $\gamma_T=\frac{1}{\sqrt{1-{\beta_T}^2}}$ is the
Lorentz factor of the transverse boost.

\section{Results and Discussion}

The correlation functions of pion and kaons were obtained for six centrality classes. For 
each centrality class, the correlation functions were obtained  in the  $k_T$ ranges 
of (0.1-0.2) , (0.2-0.3), (0.3-0.4), (0.4-0.5), (0.5-0.6), (0.6-0.7) and (0.7-0.8)  ( in GeV/c) for pion-pion pairs 
while for kaon pairs, the $k_T$ ranges started from 0.3 GeV/c  due to low statistics of kaon pairs in low $k_{T}$ bins.

\par
\begin{figure}[ht]
\begin{center}
\includegraphics[angle=0,width=0.45 \textwidth]{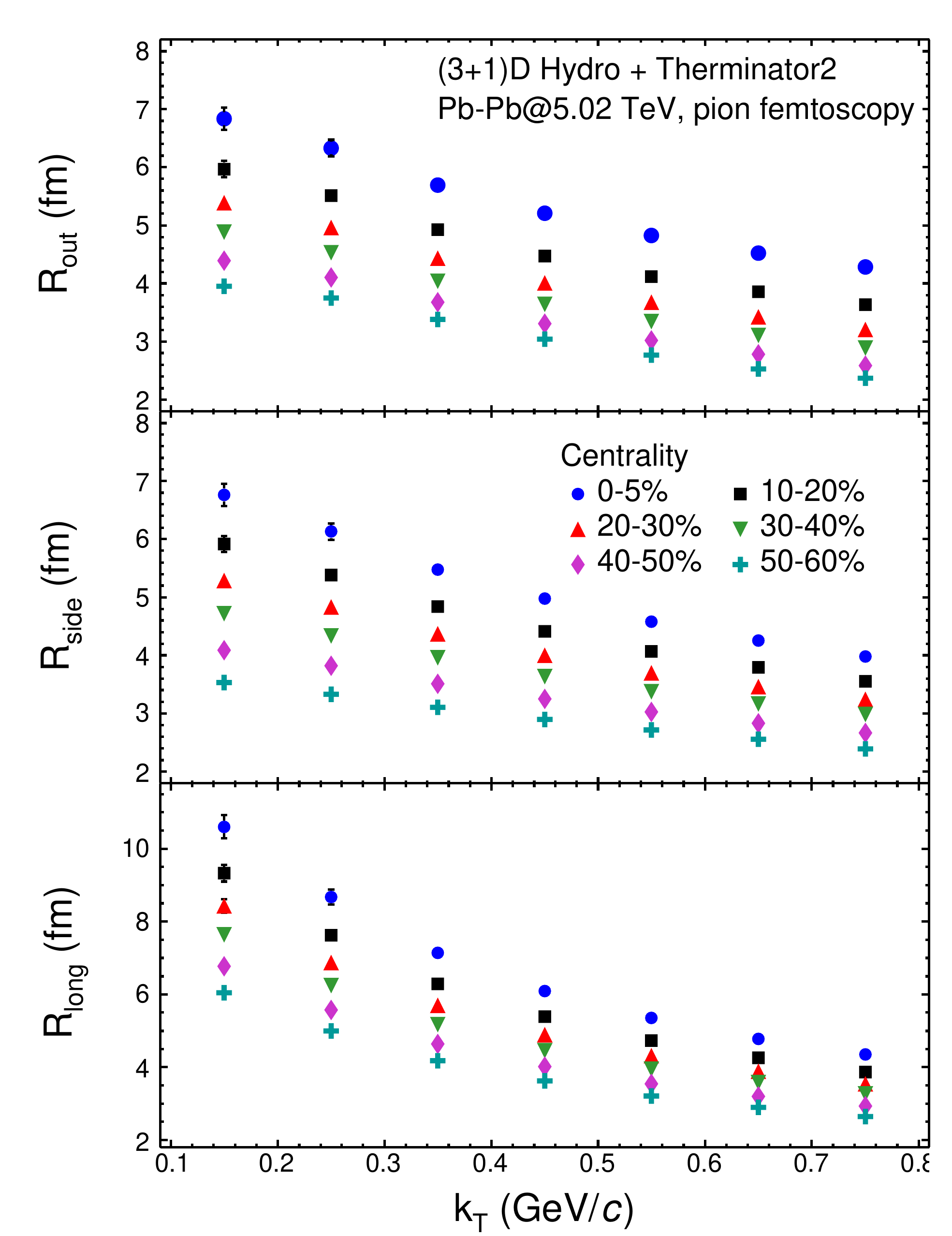}
\end{center}
\vspace{-6.5mm}
\caption{(Color online) The femtoscopic radii for pions in LCMS as a function of pair transverse momentum for different centralities
for Pb$-$Pb collisions at \rootsNN = 5.02 TeV using Therminator 2 model. }
\label{fig:pion_kT}
\end{figure}

 In Figure \ref{fig:pion_kT}, the radii for pions in LCMS are shown as a function of $k_T$ for different centrality classes
in three directions, transverse (out), side and longitudinal (long). The radii in all directions decrease with an increase in $k_{T}$ for different centrality classes. 
The maximum values of radii in both the transverse and side direction, $R_{out}$ and $R_{side}$ are 
observed to be of the order of 7 $fm$ for the lowest $k_T$ bin for most central (0-5\%) collisions. The maximum value of the radius 
along the longitudinal direction, $R_{long}$ in most central collisions is of the order of  11 $fm$ for the lowest $k_T$ range. 
The lowest radii in all three directions are found to be $\sim$ 2 $fm$, for highest $k_T$ range and for 50-60\% centrality. 
\par
\begin{figure}[ht]
\begin{center}
\includegraphics[angle=0,width=0.45 \textwidth]{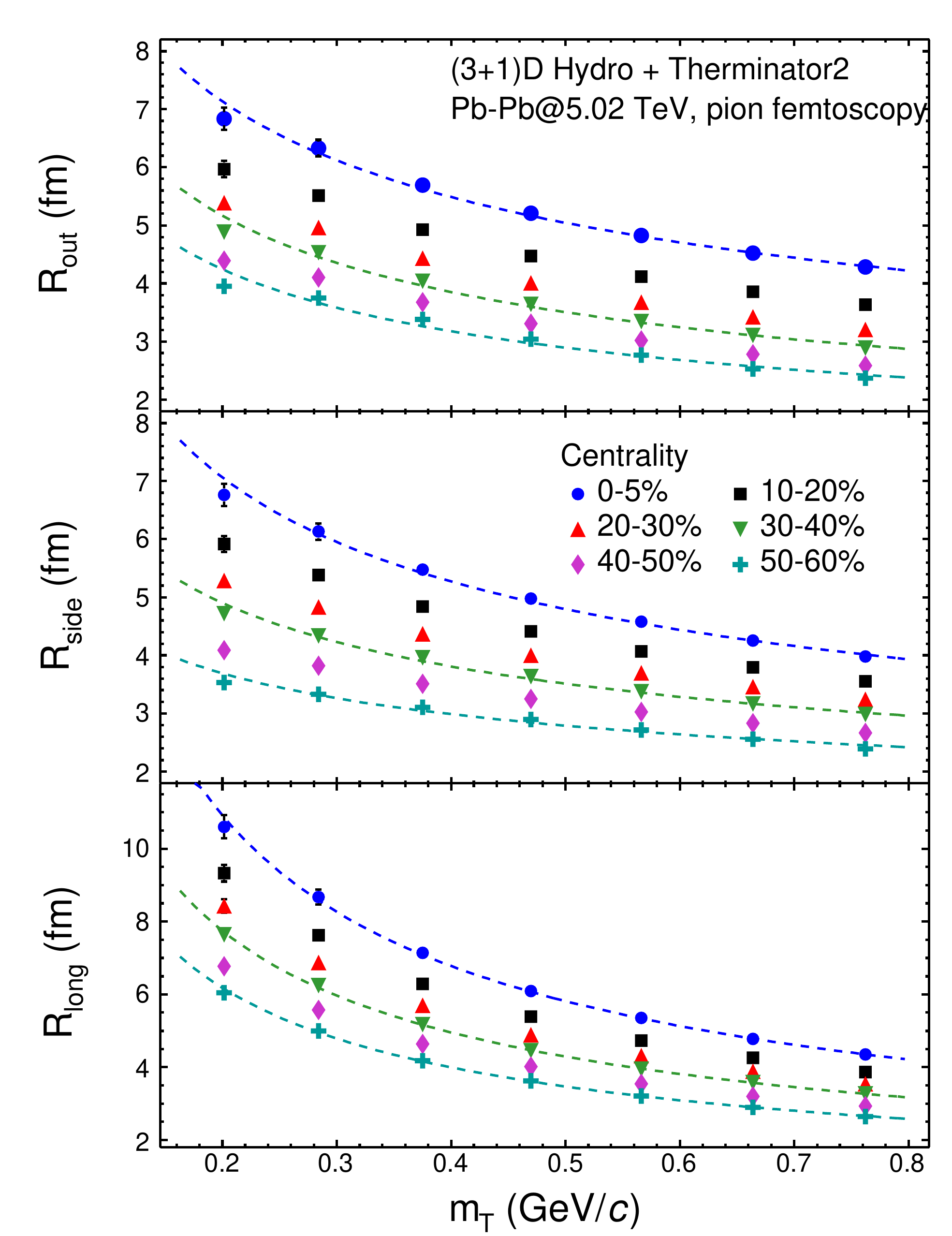}
\end{center}
\vspace{-6.5mm}
\caption{ (Color online)  The femtoscopic radii for pions in LCMS as a function of pair transverse mass at different centralities
for Pb$-$Pb collisions at \rootsNN = 5.02 TeV using Therminator 2 model. }
\label{fig:pion_mT}
\end{figure}

 In Figure \ref{fig:pion_mT}, the radii for pions in LCMS are shown as a function of transverse mass, $m_T$ for different centralities. The radii decrease as a function of  $m_T$  for all centralities. The dashed lines shown for few centralities are power-law fits  to the radii and it can be seen that the variation of radii with $m_{T}$ is well described by the power-law function. The functional form of the power law used is given by\cite{Therminator:2760MeV}:
\begin{eqnarray}
 f(m_T) = A \times m^B_{T}, \label{eq:fit_power}
\end{eqnarray}
where $A$ and $B$ are the free parameters.  The values of the extracted fits are shown in Table \ref{Table:mT_scaling_pion}. As can be seen from the table, for $out$ direction, and $side$ direction, $B$ is of similar value  for all centrality classes while it is higher for the $long$ direction, 

\begin{table}[]
\begin{tabular}{|c|c|c|c|}
\hline
Direction                      & Centrality (\%) & A           & B            \\ \hline
\multirow{3}{*}{\textit{out}}  & 0-5             & 3.88 $\pm$ 0.04 & -0.38 $\pm$
0.02 \\ \cline{2-4} 
                               & 30-40           & 2.61 $\pm$ 0.02 & -0.42 $\pm$
0.01 \\ \cline{2-4} 
                               & 50-60           & 2.17 $\pm$ 0.02 & -0.42 $\pm$
0.01 \\ \hline
\multirow{3}{*}{\textit{side}} & 0-5             & 3.58 $\pm$ 0.04 & -0.42 $\pm$
0.02 \\ \cline{2-4} 
                               & 30-40           & 2.73 $\pm$ 0.02 & -0.30 $\pm$
0.01 \\ \cline{2-4} 
                               & 50-60           & 2.26 $\pm$ 0.02 & -0.30 $\pm$
0.01 \\ \hline
\multirow{3}{*}{\textit{long}} & 0-5             & 3.61 $\pm$ 0.04 & -0.69 $\pm$
0.01 \\ \cline{2-4} 
                               & 30-40           & 2.74 $\pm$ 0.02 & -0.65 $\pm$
0.01 \\ \cline{2-4} 
                               & 50-60           & 2.24 $\pm$ 0.02 & -0.63 $\pm$
0.01 \\ \hline
\end{tabular}
\caption{The parameters obtained from the power-law fit of the radii along $out$,
$side$ and $long$ direction as a function of $m_T$ for pion-femtoscopy for selected
centralities 
for Pb$-$Pb collisions at \rootsNN = 5.02 TeV using Therminator 2 model}
\label{Table:mT_scaling_pion}
\end{table}

\par
\begin{figure}[ht]
\begin{center}
\includegraphics[angle=0,width=0.45 \textwidth]{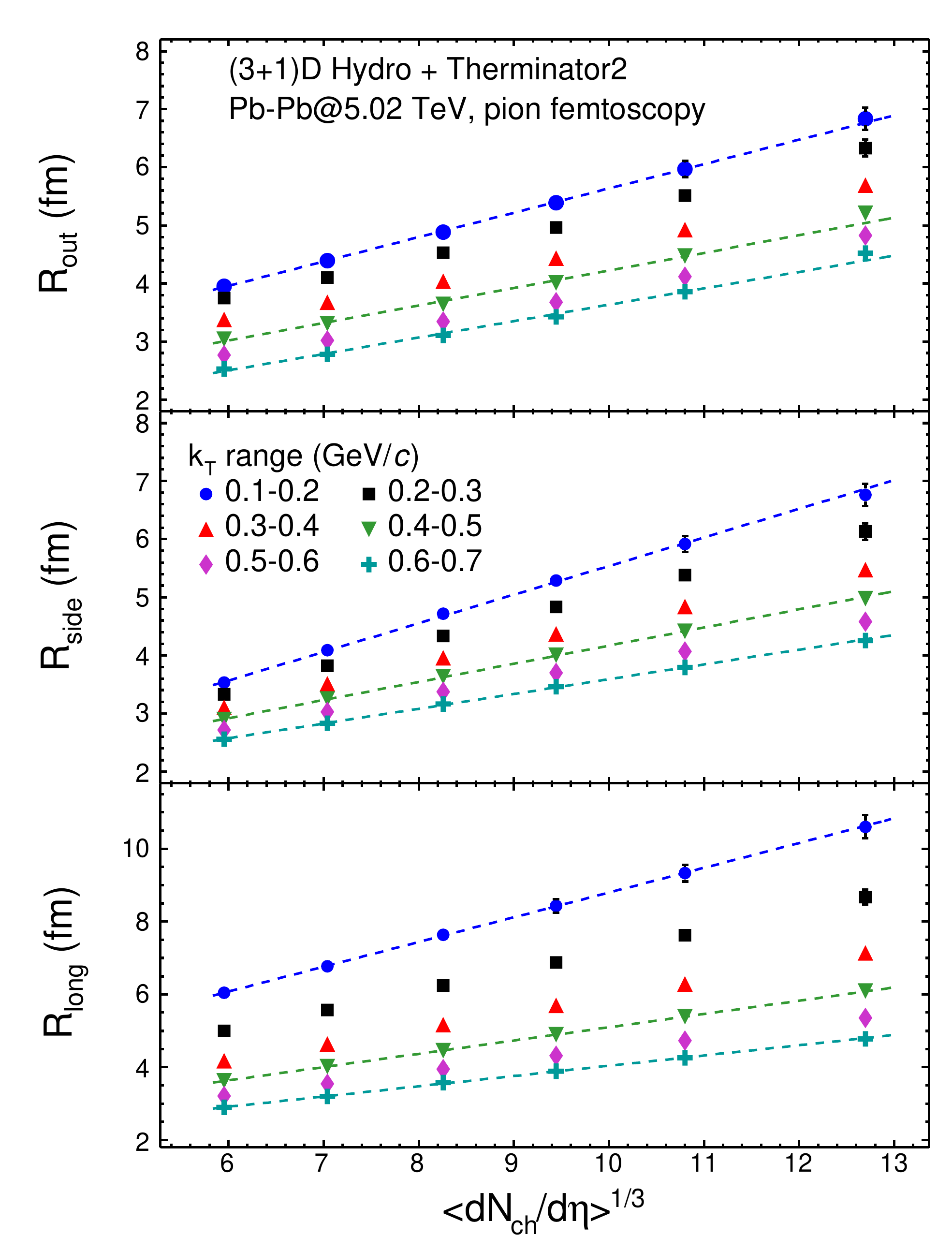}
\end{center}
\vspace{-6.5mm}
\caption{(Color online) The femtoscopic radii of pions as a function $\left<\frac{dN_{ch}}{d\eta}\right>^\frac{1}{3}$  for several $k_T$ ranges
for Pb$-$Pb collisions at \rootsNN = 5.02 TeV using Therminator 2 model}
\label{fig:pion_dNdeta}
\end{figure}

 In Fig. \ref{fig:pion_dNdeta}, the dependence of radii of pions on the final-state multiplicity for different $k_T$ ranges has been shown. 
It is observed that the radii in all directions decrease from most central events to peripheral events. The values are fitted with  a linear function
given by:\\
\begin{eqnarray}
 f( \left<\frac{dN_{ch}}{d\eta}\right>^\frac{1}{3}  ) = C + D \times \left<\frac{dN_{ch}}{d\eta}\right>^\frac{1}{3}, \label{eq:fit_line}
\end{eqnarray}
for some selected $k_T$ ranges. The values of the extracted parameters are shown in Table \ref{Table:dNdeta_scaling_pion}. In all three directions,  D decreases  for higher 
$k_T$ regions. 
\begin{table}[]
\begin{tabular}{|c|c|c|c|}
\hline
Direction                      & $k_{T}$  (GeV/\textit{c}) & C          &  D     
    \\ \hline
\multirow{3}{*}{\textit{out}}  & 0.1-0.2            & 1.44 $\pm$ 0.15 & 0.42 $\pm$
0.02 \\ \cline{2-4} 
                               & 0.4-0.5            & 1.20 $\pm$ 0.07 & 0.30 $\pm$
0.01 \\ \cline{2-4} 
                               & 0.6-0.7            & 0.81 $\pm$ 0.05 & 0.28 $\pm$
0.01 \\ \hline
\multirow{3}{*}{\textit{side}} & 0.1-0.2            & 0.61 $\pm$ 0.14 & 0.49 $\pm$
0.02 \\ \cline{2-4} 
                               & 0.4-0.5            & 1.04 $\pm$ 0.07 & 0.31 $\pm$
0.01 \\ \cline{2-4} 
                               & 0.6-0.7            & 1.04 $\pm$ 0.05 & 0.25 $\pm$
0.01 \\ \hline
\multirow{3}{*}{\textit{long}} & 0.1-0.2            & 1.99 $\pm$ 0.25 & 0.68 $\pm$
0.03 \\ \cline{2-4} 
                               & 0.4-0.5            & 1.45 $\pm$ 0.09 & 0.36 $\pm$
0.01 \\ \cline{2-4} 
                               & 0.6-0.7            & 1.22 $\pm$ 0.06 & 0.28 $\pm$
0.01 \\ \hline
\end{tabular}
\caption{The parameters obtained from the linear fit of the radii along $out$,
$side$ and $long$ direction as a function of
$\left<dN_{ch}/d\eta\right>^{\frac{1}{3}}$ for pion-femtoscopy 
for selected centralities for Pb$-$Pb collisions at \rootsNN = 5.02 TeV using
Therminator 2 model} \label{Table:dNdeta_scaling_pion}
\end{table}

\par The scaling of $m_T$ and $(\frac{dN_{ch}}{d\eta})^{1/3}$ shows the presence of collective flow in both the transverse dimensions \cite{Therminator:2760MeV}.

\par
\begin{figure}[ht]
\begin{center}
\includegraphics[angle=0,width=0.45 \textwidth]{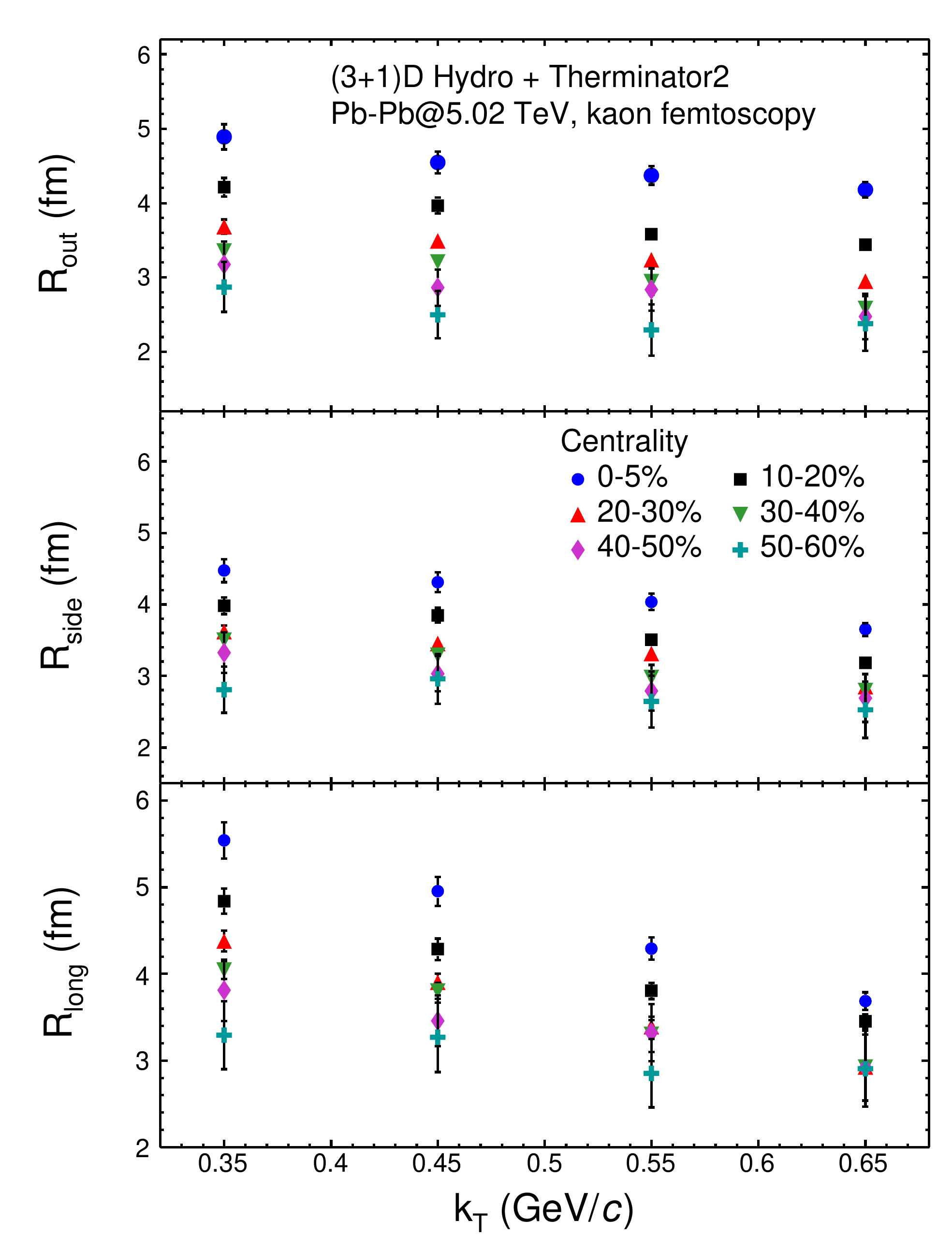}
\end{center}
\vspace{-6.5mm}
\caption{(Color online) The femtoscopic radii for kaons in LCMS as a function of $m_{T}$ for different centralities
for Pb$-$Pb collisions at \rootsNN = 5.02 TeV using Therminator model 2. }
\label{fig:kaon_kT}
\end{figure}

\begin{figure}[ht]
\begin{center}
\includegraphics[angle=0,width=0.45 \textwidth]{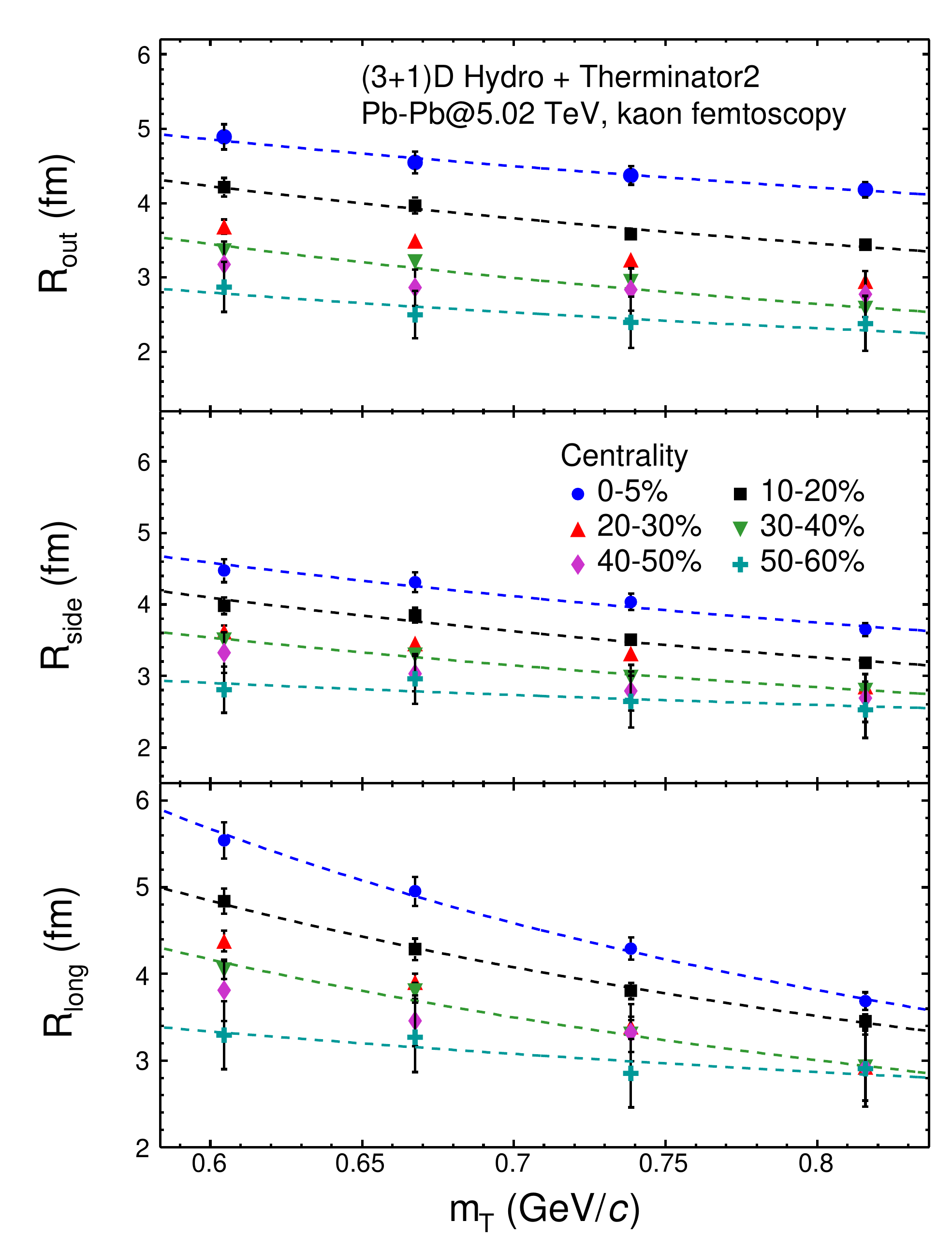}
\end{center}
\vspace{-6.5mm}
\caption{ (Color online)  The femtoscopic radii for kaons in LCMS as a function of pair transverse mass at different centralities
for Pb$-$Pb collisions at \rootsNN = 5.02 TeV using Therminator 2 model. }
\label{fig:kaon_mT}
\end{figure}

\begin{figure}[ht]
\begin{center}
\includegraphics[angle=0,width=0.45 \textwidth]{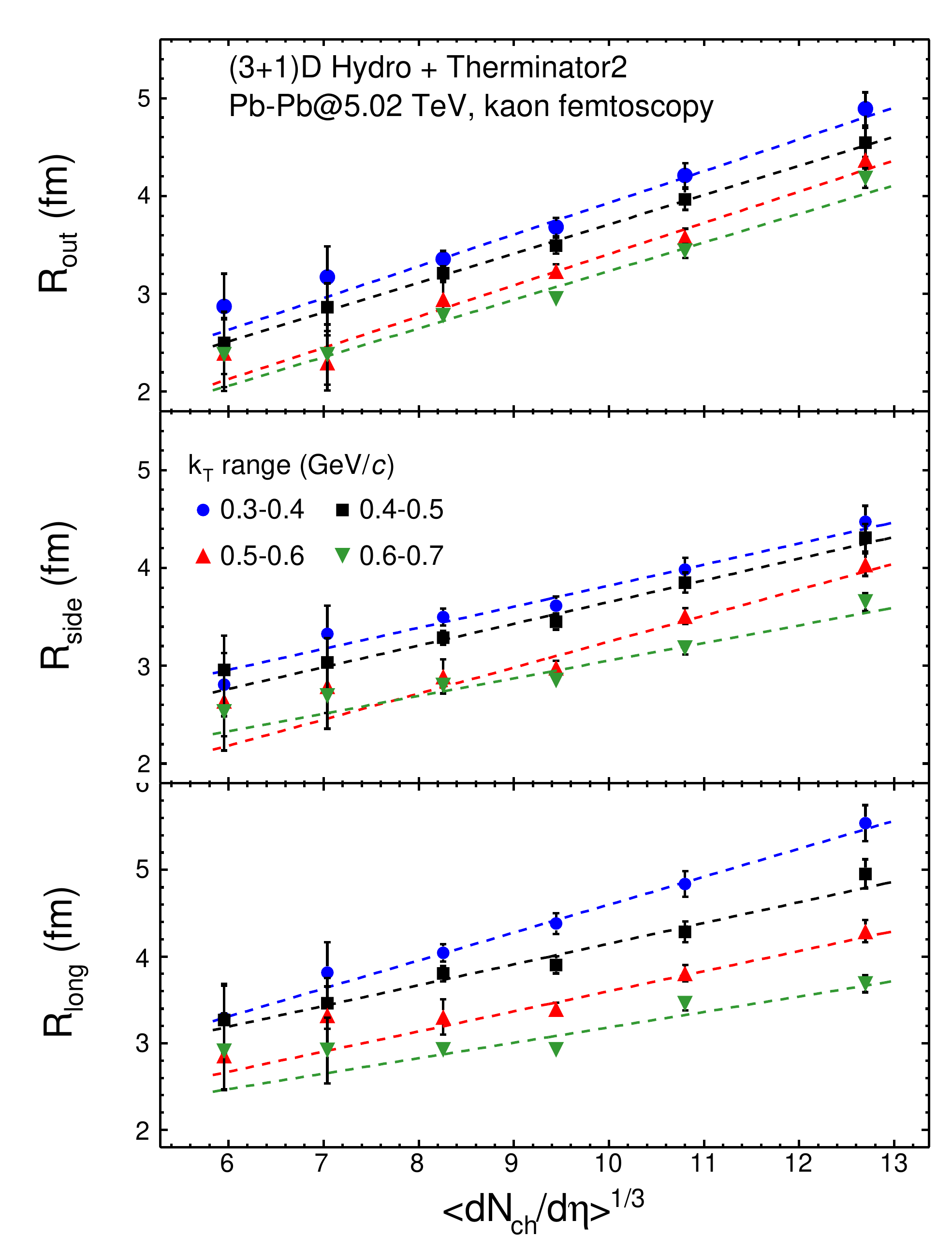}
\end{center}
\vspace{-6.5mm}
\caption{(Color online) The femtoscopic radii for kaons as a function of cube root of the charged particle multiplicity for several $k_T$ ranges
for Pb$-$Pb collisions at \rootsNN = 5.02 TeV using Therminator 2 model}
\label{fig:kaon_dNdeta}
\end{figure}

 The variation of radii for kaons in LCMS as a function of  $k_T$ and $m_T$  for different centralities are shown in Fig. \ref{fig:kaon_kT} and Fig. 
\ref{fig:kaon_mT}, respectively. The radii decrease with $k_T$  and $m_T$ for the considered centrality  classes and also from most central to 
peripheral events. The trend is similar to the one observed for pions. The dashed lines in Figure \ref{fig:kaon_mT} depict the power law fits (given by Eq. \ref{eq:fit_power}.) to the radii values. The extracted fit parameters are shown in Table \ref{Table:mT_scaling_kaon}. 
Figure \ref{fig:kaon_dNdeta}, shows the final-state multiplicity dependence of the radii for kaons in different $k_T$ ranges. The radii increases monotonically from peripheral 
to central  events for different $k_{T}$ ranges in all three directions. The statistical uncertainties in the radii at
lower multiplicity region is relatively larger in all directions. The values  have been fitted with  Eq. \ref{eq:fit_line} for some selected $k_T$ ranges and the values are 
shown in Table \ref{Table:dNdeta_scaling_kaon}. 

\begin{table}[]
\begin{tabular}{|c|c|c|c|}
\hline
Direction                      & Centrality (\%) & A           & B            \\ \hline
\multirow{4}{*}{\textit{out}}  & 0-5             & 3.77 $\pm$ 0.17 & -0.50 $\pm$
0.13 \\ \cline{2-4} 
                               & 10-20           & 2.96 $\pm$ 0.11 & -0.70 $\pm$
0.11  \\ \cline{2-4} 
                               & 30-40           & 2.15 $\pm$ 0.07 & -0.75 $\pm$
0.09 \\ \cline{2-4} 
                               & 50-60           & 2.00 $\pm$ 0.50 & -0.65 $\pm$
0.63 \\ \hline
\multirow{4}{*}{\textit{side}} & 0-5             & 3.20 $\pm$ 0.15 & -0.70 $\pm$
0.13 \\ \cline{2-4} 
                               & 10-20           & 2.73 $\pm$ 0.10 & -0.79 $\pm$
0.11 \\ \cline{2-4} 
                               & 30-40           & 2.40 $\pm$ 0.08 & -0.76 $\pm$
0.09 \\ \cline{2-4} 
                               & 50-60           & 2.37 $\pm$ 0.54 & -0.39 $\pm$
0.58 \\ \hline
\multirow{4}{*}{\textit{long}} & 0-5             & 2.80 $\pm$ 0.14 & -1.38 $\pm$
0.14 \\ \cline{2-4} 
                               & 10-20           & 2.74 $\pm$ 0.11 & -1.11 $\pm$
0.12 \\ \cline{2-4} 
                               & 30-40           & 2.33 $\pm$ 0.07  & -1.13 $\pm$
0.09 \\ \cline{2-4} 
                               & 50-60           & 2.55 $\pm$ 0.60 & -0.52 $\pm$
0.60 \\ \hline
\end{tabular}
\caption{The parameters obtained from the power-law fit of the radii along $out$,
$side$ and $long$ direction as a function of $m_T$ for kaon-femtoscopy for selected
centralities 
        for Pb$-$Pb collisions at \rootsNN = 5.02 TeV using Therminator 2 model}
\label{Table:mT_scaling_kaon}
\end{table}

\begin{table}[]
\begin{tabular}{|c|c|c|c|}
\hline
Direction                      & $k_{T}$ (GeV/c) & C          & D           \\
\hline
\multirow{4}{*}{\textit{out}}  & 0.3-0.4            & 0.69 $\pm$ 0.32 & 0.32 $\pm$
0.03 \\ \cline{2-4} 
                               & 0.4-0.5            & 0.72 $\pm$ 0.28 & 0.30 $\pm$
0.03 \\ \cline{2-4} 
                               & 0.5-0.6            & 0.21 $\pm$ 0.33 & 0.32 $\pm$
0.03 \\ \cline{2-4} 
                               & 0.6-0.7            & 0.30 $\pm$ 0.20 & 0.29 $\pm$
0.02 \\ \hline
\multirow{4}{*}{\textit{side}} & 0.3-0.4            & 1.66 $\pm$ 0.32 & 0.22 $\pm$
0.03 \\ \cline{2-4} 
                               & 0.4-0.5            & 1.43 $\pm$ 0.28 & 0.22 $\pm$
0.03 \\ \cline{2-4} 
                               & 0.5-0.6            & 0.59 $\pm$ 0.32 & 0.27 $\pm$
0.03 \\ \cline{2-4} 
                               & 0.6-0.7            & 1.25 $\pm$ 0.20 & 0.18 $\pm$
0.02 \\ \hline
\multirow{4}{*}{\textit{long}} & 0.3-0.4            & 1.37 $\pm$ 0.39 & 0.32 $\pm$
0.04 \\ \cline{2-4} 
                               & 0.4-0.5            & 1.75 $\pm$ 0.33 & 0.24 $\pm$
0.04 \\ \cline{2-4} 
                               & 0.5-0.6            & 1.28 $\pm$ 0.36 & 0.23 $\pm$
0.03 \\ \cline{2-4} 
                               & 0.6-0.7            & 1.40 $\pm$ 0.21 & 0.18 $\pm$
0.02 \\ \hline
\end{tabular}
\caption{The parameters obtained from the linear fit of the radii along $out$,
$side$ and $long$ direction as a function of
$\left<dN_{ch}/d\eta\right>^{\frac{1}{3}}$ for kaon pairs 
        for selected centralities for Pb$-$Pb collisions at \rootsNN = 5.02 TeV using
Therminator 2 model} \label{Table:dNdeta_scaling_kaon}
\end{table}

\begin{figure}[ht]
\begin{center}
\includegraphics[angle=0,width=0.45 \textwidth]{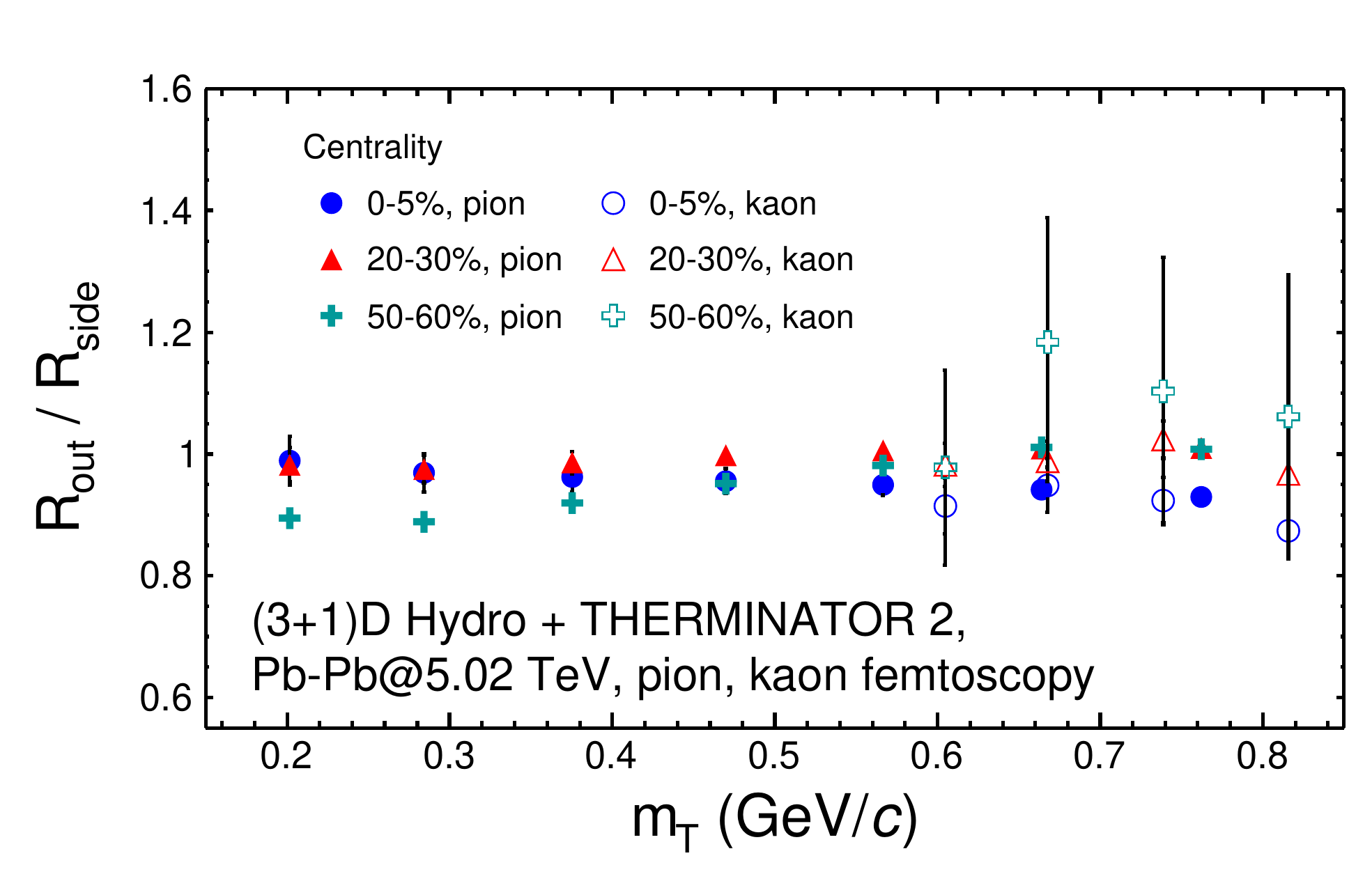}
\end{center}
\vspace{-6.5mm}
\caption{ (Color online)  $R_{out}$/$R_{side}$  as a function of  $m_{T}$  for pions and kaons  for different  centrality classes for Pb$-$Pb collisions at \rootsNN = 5.02 TeV using Therminator 2 model. }
\label{fig:pion_kaon_ratio}
\end{figure}

One can observe that the slopes  of $m_{T}$ dependence of kaon radii are quite different from the pions.  To test the hypothesis of common effective scaling for $m_T$, the pion and kaon radii have been plotted simultaneously as a function of $m_T$ for some selected
centralities and they have been fitted with the power-law function of $m_T$ given by Eq. \ref{eq:fit_power}. The fit parameters are that of pions.  It can be seen from the figure that the  $m_{T}$ scaling is approximately followed in  $out$ and $side$ direction while it is slightly violated for (0-20)\% and (20-30)\% centrality in $long$ direction. 
From these observations, it is clear that the (3+1)D Hydro + THERMINATOR 2 model predicts an approximate scaling of the three-dimensional femtoscopic radii as function 
of pair $m_T$ in LCMS for pions and kaons which has also been claimed in the femtoscopic study of identical particles produced in Pb$-$Pb collision at $\sqrt{s_{NN}}=2.76$ 
TeV modeled in (3+1)D hydrodynamics \cite{Therminator:2760MeV}. The femtoscopic radii for protons in LCMS could not be included in this observation due to the statistical 
limitations but from this scaling, we can predict the radii of not only protons but heavier particles also.  
Both $R_{out}$ and $R_{side}$ radii are influenced by flow and re-scatterings, and thus  their ratio becomes a robust observable which is not affected by these effects.  
Figure \ref{fig:pion_kaon_ratio}  also shows the $m_{T}$ dependence of the ratio  $R_{out}/R_{side}$ for pions and kaons.  It can be seen that the values for kaons are slightly 
lower than that of pions for (0-5)\% centrality while it is mildly higher than pions for peripheral collisions. However, the statistical uncertainties are large and one cannot make any conclusions
about the different space-time correlations for these two particle species.

\begin{figure}[ht]
\begin{center}
\includegraphics[angle=0,width=0.45 \textwidth]{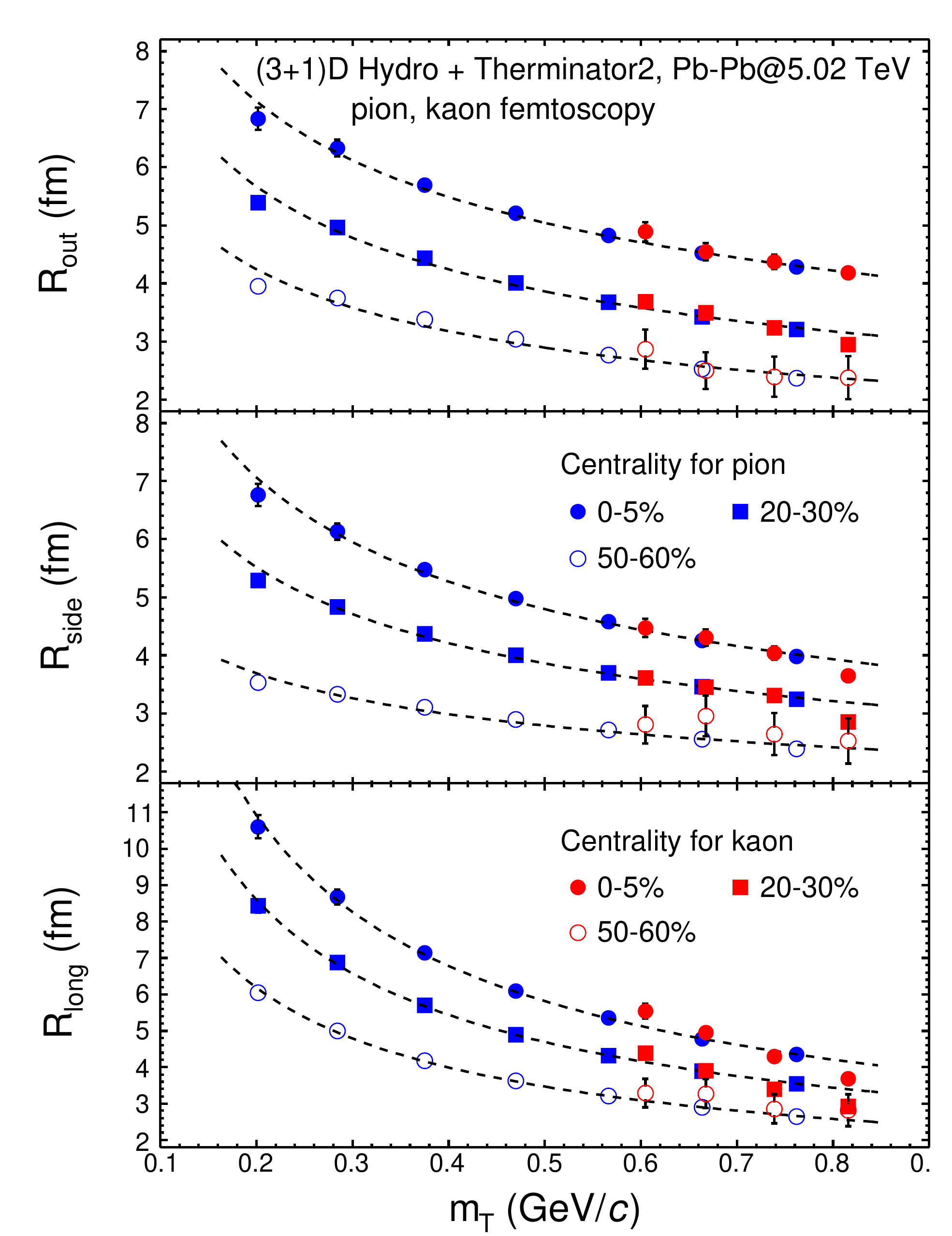}
\end{center}
\vspace{-6.5mm}
\caption{ (Color online) The femtoscopic radii in LCMS for pions and
kaons for selected centralities for Pb$-$Pb collisions at \rootsNN = 5.02 TeV using Therminator 2 model. Lines represent power-law 
fits to the combined pion and kaon data points at a given centrality and direction. }
\label{fig:pion_kaon_mT}
\end{figure}

\par
\begin{figure}[ht]
\begin{center}
\includegraphics[angle=0,width=0.45 \textwidth]{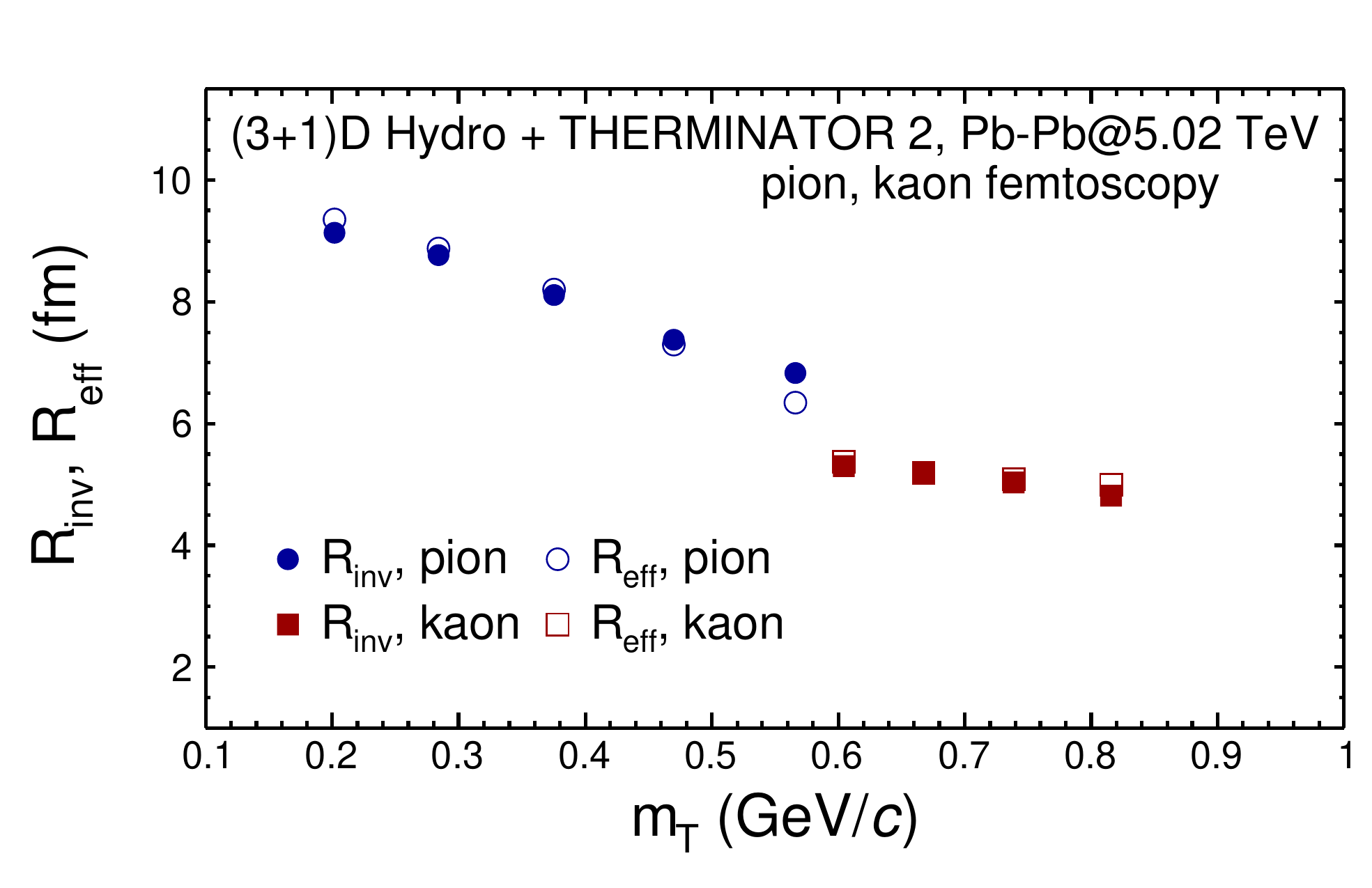}
\end{center}
\vspace{-6.5mm}
\caption{ (Color online) Comparison of the $R_{inv}$ obtained directly from the fit to the one-dimensional correlation function in PRF and a result $R_{eff}$ of 
the approximate procedure to estimate R inv from values of $R_{out}$, $R_{side}$ and $R_{long}$ measured in LCMS}
\label{fig:Rinv_Reff}
\end{figure}

The extraction of femtoscopic radii in three dimensions and in different centrality classes for heavier particles like kaons and protons are sometimes
limited by statistics. Therefore,  a measurement of one-dimensional radius in the PRF (pair rest frame) is often performed. The $R_{inv}$ is a 
direction-averaged source size in PRF and can be obtained by fitting the one-dimensional correlation function with Eq. \ref{eq:fit1d} in 
PRF \cite{Therminator:2760MeV} which assumes that the source is Gaussian. 
The system in LCMS can be Lorentz boosted in the direction of pair transverse momentum to PRF to obtain the radii $R^{*}_{out}$, $R^{*}_{side}$ and $R^{*}_{long }$  in the PRF.  An effective radius, $R_{eff}$, can be calculated by taking the mean of $R^{*}_{out}$, $R^{*}_{side}$ and $R^{*}_{long}$ 
and can be compared to $R_{inv}$. In Fig. \ref{fig:Rinv_Reff}, the $R_{inv}$ for pions
and kaons (obtained from a fit of Eq. \ref{eq:fit1d} to the one-dimensional correlation function) have been compared with the $R_{eff}$ calculated from the corresponding $R_{out}$, $R_{side}$ and $R_{long}$ obtained by fitting three dimensional 
correlation function in LCMS and also the corresponding $\gamma_T$. One can observe that there is a good agreement between the $R_{inv}$ and the corresponding $R_{eff}$ and the differences are within 7\%.This indicates that once can determine $R_{inv}$ from the values of $R_{out}$, $R_{side}$ and $R_{long}$ obtained in LCMS and from the Lorentz boost factor.

\par
\begin{figure}[ht]
\begin{center}
\includegraphics[angle=0,width=0.45 \textwidth]{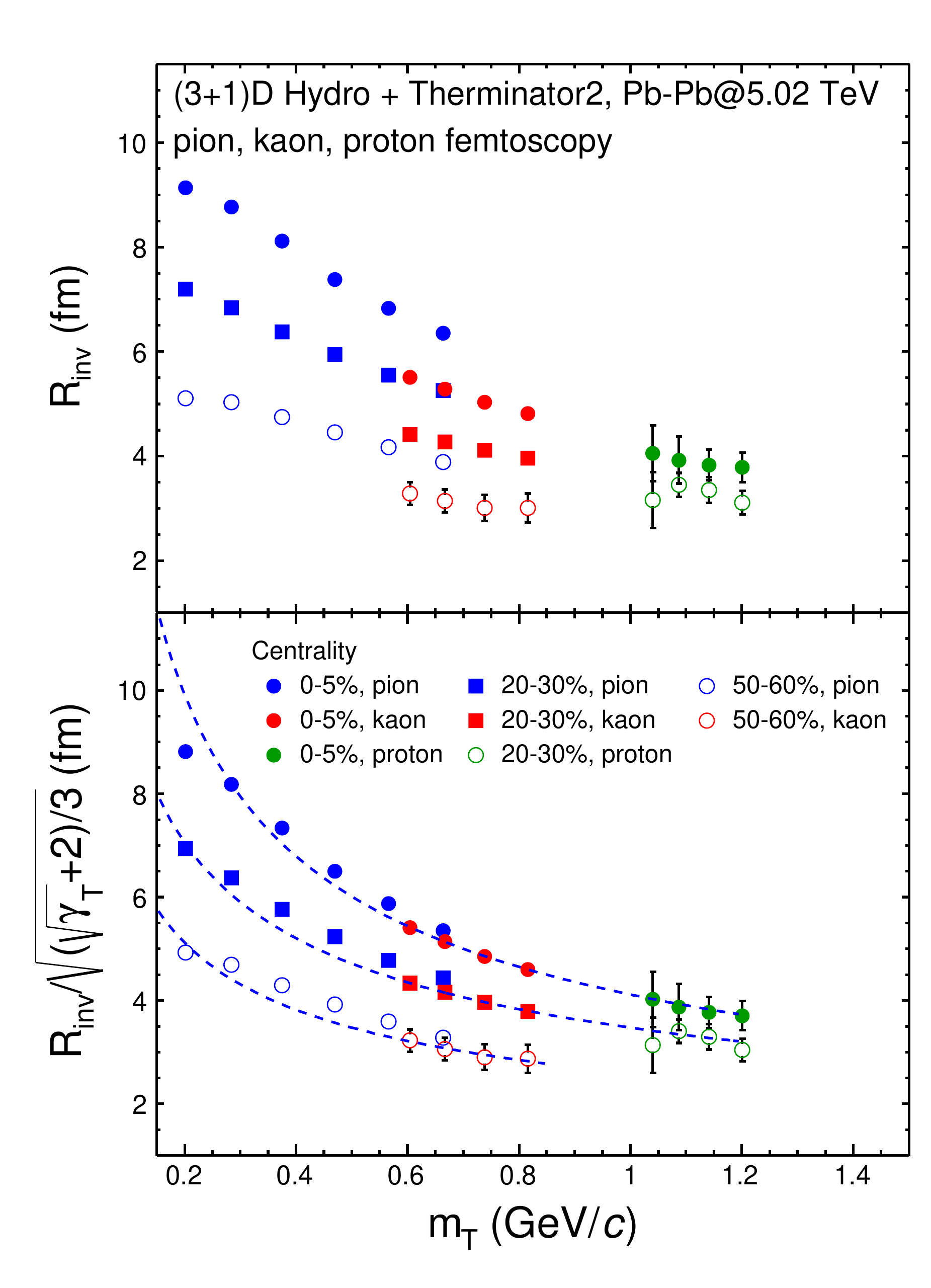}
\end{center}
\vspace{-6.5mm}
\caption{ (Color online) a) One-dimensional femtoscopic radius
$R_{inv}$ for pions, kaons and protons calculated in the PRF for selected
centralities. (b) R inv for pions, kaons and protons scaled with the
kinematic factor, for selected centralities (see text for details). Lines
represent power-law fits}
\label{fig:Rinv_mT_all}
\end{figure}

 Due to statistical limitations  of performing the 3-D femtoscopic studies of proton,  the one-dimensional radius $R_{inv}$ in PRF for protons  was obtained with this model. 
As a common effective $m_T$ scaling have been observed for radii of pion and kaon in 
LCMS and $R_{inv}$  was shown to be directly related to them, one expects to observe a similar scaling of $R_{inv}$ for pions, kaons and protons also. However, for the same $m_T$, $\gamma_T$ would be different for pions and kaons and the difference would be more for protons also. This would result in different  $R^*_{out}$ values for pions, kaons and
protons for the same $m_T$ bin. Therefore $R_{inv}$ of pions, kaons and protons are expected to differ for same $m_T$. Fig. \ref{fig:Rinv_mT_all} shows the  $R_{inv}$ for pions, kaons and protons in PRF as a function of $m_T$ for different centralities classes.  It can be clearly seen in the upper panel of the figure that pions, kaons and protons show different trends for different centralities at similar $m_{T}$.

As discussed previously,  the violation of scaling in $R_{inv}$ can be attributed to difference in Lorentz boost factors for different particles. The 
 one dimensional radii therefore can be scaled by the following factor 
 \begin{eqnarray}
f = (\gamma_T  + 2)/3, \label{eq:scale_gamma} 
\end{eqnarray}
It can be seen that a common scaling for pions, kaons and protons is observed and is seen in lower panel of Fig. \ref{fig:Rinv_mT_all}. The form of 
this scaling factor is described in \cite{Therminator:2760MeV}. Thus, the measurement of the one-dimensional radius of different particle species 
in PRF can give important information for search of collectivity scaling in experimental data analysis.

\section{Conclusion}
The femtoscopic radii extracted from system of  identical pions, kaons and protons using (3+1)D hydrodynamic model coupled with THERMINATOR 2 code for statistical hadronization, resonance propagation  and decay has been studied as a function of centrality and  pair transverse mass, $m_{T}$.  The effect of hadronic re-scattering has not been considered in this model. The obtained radii decreased from central to peripheral collisions as well as with an increase 
of  $m_T$. In the LCMS frame, the radii followed two mutually exclusive effective scaling, a power law scaling with pair $m_T$ and a linear scaling with $\left<dN_{ch}/d\eta \right>^\frac{1}{3}$. The slope for $\left<dN_{ch}/d\eta \right>^\frac{1}{3}$  scaling was observed to be  almost similar in all the three directions ($out$, $side$ and $long$). For the $m_T$ scaling, the curves  in $out$ and $side$ directions were less steeper compared to  $long$ direction which corresponds to the larger flow velocity \cite{scaling1}. A violation of such scaling was observed recently in  LHC data for  kaons and in other hydrodynamical models which incorporated the effects of 
hadronic re-scattering. Therefore, violation of such scalings in experimental data probes the relevance of hadronic re-scattering phase.
The one-dimensional femtoscopic radii $R_{inv}$ in PRF have also been studied as a function of centrality and pair $m_T$. A 
power-law $m_T$ scaling, similar to the three dimensional radii measured in LCMS, have also been observed for R$_{inv}$ though the scalings 
in LCMS and PRF were independent to each other. A common effective scaling of $R_{inv}$  for pions, kaons and protons was observed after scaling the 
obtained $R_{inv}$ with a kinematic factor arising due to difference in Lorentz factors for different species of particles.

\section{Acknowledgements}
The authors would like to thank Department of Science 
and Technology (DST), Government of India for supporting the present 
work. The authors would also like to thank Piotr Bozek for providing the files with hypersurfaces from the
(3+1) dimensional perfect hydrodynamics.

\end{document}